\documentclass[twocolumn,showpacs,prb,superscriptaddress]{revtex4}

\usepackage{graphicx}
\usepackage{subfigure}
\usepackage{dcolumn}
\usepackage{bm}

\newcommand{\mbf}{\mathbf}

\begin{document}

\title{Fractional periodicity and magnetism of extended quantum rings}

\author{Y. Hancock}
\email{yvh@fyslab.hut.fi}
\affiliation{Laboratory of Physics, Helsinki University of Technology, PO Box 4100, FI-02015 HUT, Finland}
\author{J. Suorsa}
\affiliation{Laboratory of Physics, Helsinki University of Technology, PO Box 4100, FI-02015 HUT, Finland}
\author{E. T\"ol\"o}
\affiliation{Laboratory of Physics, Helsinki University of Technology, PO Box 4100, FI-02015 HUT, Finland}
\affiliation{Helsinki Institute of Physics, PO Box 64, FI-00014, University of Helsinki, Finland}
\author{A. Harju}
\affiliation{Laboratory of Physics, Helsinki University of Technology, PO Box 4100, FI-02015 HUT, Finland}
\affiliation{Helsinki Institute of Physics, PO Box 64, FI-00014, University of Helsinki, Finland}

\date{\today}

\begin{abstract} 
The magnetic properties and nature of the persistent current in small
flux-penetrated $t-t'-U$ rings are investigated. An effective
rigid-rotator description is formulated for this system, which
coincides with a transition to a ferromagnetic state in the model. The
criteria for the onset of effective rigid rotation is given. The model
is used to understand continuum model ground-state solutions for a
$2D$ few-particle hard-wall quantum dot, where ferromagnetic solutions are found even without the Zeeman
coupling to spin. After the onset of effective
rigid rotation, a $97$--$98\%$ correspondence can be determined
between the lattice model and continuum model eigenstate results.
\end{abstract}

\maketitle

\section{\label{sec:level1}INTRODUCTION\protect\\ }
Electron traps in semi-conductor materials\cite{smr1} are nanosystems
that are of intense current interest. Examples include quantum rings
and dots, which, due to their atom-like features, have shown immense
potential for technological application. Recently, it has been a goal
to acquire a more fundamental understanding of these systems. For
quantum rings, this interest has been partly motivated by the
experimental observation of Aharonov--Bohm oscillations and persistent
currents.\cite{al1,ufk1} In the strictly $1D$ limit, the quantum ring
can be studied by applying both a discrete model, such as the Hubbard
model,\cite{jh1} and continuum model approaches. In the strongly
interacting limit, a correspondence is found between these two sets of
results---the electrons become localized and the spin-state of the
system becomes that of the anti-ferromagnetic Heisenberg
Hamiltonian.\cite{sv1,mk1} Electron localization causes the system to
become a `rigid-rotator'. In this case, only the rotational degree of
freedom exists so that a change in the magnetic flux results in a
change in the angular momentum state of the ring.

The fractional periodicity seen in the Hubbard model and continuum
model solutions has been well documented.\cite{ny1,fvk1,fvk2,fvk3}
Physical insight into these properties can be obtained from the
ground-state energy Bethe Ansatz results for the $1D$ Hubbard chain
with magnetic flux.\cite{ny1} The ground-state energy as a function of
flux, $\Phi$, consists of a sequence of parabolic segments having
$\Phi_{o}/N_{e}$ periodicity. Here, $\Phi_{o}=h/e$, is the flux
quantum and $N_{e}$ is the number of electrons. In the strongly
interacting limit, the ground-state solution at a given value of the
flux can be obtained by creating a single hole in the magnon sea. As
the flux is increased, the hole is found to move from one Fermi point
to the next. The $U\rightarrow\infty$ Bethe Ansatz solutions of the
Hubbard model show that the energy has a $(M -
N_{e}\frac{\Phi}{\Phi_{0}})^2$ flux dependence.\cite{sv1} The energy
minima are found at $\Phi/\Phi_{o}=M/N_{e}$, where $M$ is the angular
momentum.  The minima of the parabolas and fractional periodicity are
related to the change in angular momentum state as the flux is
increased through the ring. In the $U\rightarrow\infty$ limit, the
ground-state Bethe Ansatz solution becomes that of the
anti-ferromagnetic Heisenberg chain,\cite{ny1} thereby justifying both
the rigid-rotator description and anti-ferromagnetism seen in the
continuum model results.\cite{mk1}

The present work explores the fractional periodicity in
flux-penetrated $t-t'-U$ ring systems. The $t-t'-U$ lattice model is a
Hubbard model with an additional next-nearest-neighbor hopping, $t'$,
term.\cite{sd1} For any $1D$ system with nearest-neighbor hopping, the
ground-state is unmagnetized when there is a real and
particle-symmetric, density-dependent interaction.\cite{lieb1} Thus,
these criteria are met for the single-band Hubbard model, but not for
the $t-t'-U$ system due to the particle-hole symmetry breaking $t'$
term.\cite{sd1} In the latter case, no definite order is enforced to
the particles, and, thus, the possibility of a ferromagnetic state
arises. Studies on the $1D$, $t-t'-U$ chain have indeed verified an
extensive ferromagnetic phase for this system.\cite{sd1} Although
much work has been done to understand the $t-t'-U$ model, little is
understood about the $t-t'-U$ flux-penetrated ring.  Recent
investigations of the kinetic $t-t'$ $(U=0)$ system with magnetic-flux
penetration have indicated an order of magnitude increase in the
persistent current due to the addition of the next-nearest-neighbor
hopping term.\cite{skm1} To date, no investigations on the role of
interaction effects in flux-penetrated, extended-hopping systems have
been made.

The physics of the flux-penetrated $t-t'-U$ ring will be shown to explain the continuum model solutions for a circularly symmetric, $2D$ hard-wall quantum dot. In the quantum dot, the strong electron--electron interactions confine the electrons to a ring-like geometry of finite width. The continuum model solution for the flux-penetrated system is shown to have ferromagnetic correlations, with fractional periodicity in the ground-state energy as a function of the penetrating flux. The ferromagnetism seen in this system is in contrast to the anti-ferromagnetic ground-state result obtained in the limit of the purely $1D$ ring. The $t-t'-U$ model, which is considered to be the minimal model for ferromagnetism, is therefore thought to be an appropriate choice for probing the ferromagnetic behaviour seen in the continuum model results. By comparing the lattice model and the continuum model solutions, criteria can then be established for obtaining a correspondence between these two sets of results. These criteria can then be used to obtain a microscopic understanding of the ferromagnetism in the dot, as well as to explain the fractional periodicty in the ground-state energy results.

The paper is organized as follows. Section II describes the lattice model and theoretical method. This is followed by Section III, the results and discussion, which is divided into two parts. Part A of Section III details the quarter-filled, flux-penetrated $t-t'-U$ ring system. The flux-penetrated Hubbard model is first discussed as this provides an appropriate context by which to understand the flux-penetrated $t-t'-U$ results. Specific criteria are then established to explain the onset of fractional periodicity seen in the $t-t'-U$ ring as a function of $t'/t$ and $U/t$. Part B of Section III shows an application of the physics of the flux penetrated $t-t'-U$ ring by determining the correspondence between it and the continuum model results for a few-particle, $2D$ hard-wall quantum dot. The criteria for this correspondence are given together with a discussion of the underlying physical mechanisms which lead to these results. Section IV concludes the main findings of this work.

\section{\label{sec:level1}LATTICE MODEL \& THEORETICAL METHOD\protect\\ }
The Hamiltonian for the single-band, flux-penetrated $t-t'-U$ ring is
\begin{eqnarray}H=&-&t\sum_{i\sigma}(c^{\dag}_{i+1\sigma}c_{i\sigma}e^{-i\phi} + H.c.)\nonumber\\
&-& t'\sum_{i\sigma}(c^{\dag}_{i+2\sigma} c_{i\sigma}e^{-2i\phi} + H.c.)\nonumber\\
&+&\sum_{i}U_{i}n_{i\uparrow}n_{i\downarrow}. \label{hamiltonian}
\end{eqnarray}
Here, $c^{\dag}_{i\sigma}$ $(c_{i\sigma})$ creates (destroys) an
electron with spin $\sigma=\{\uparrow,\downarrow\}$ at site $i$ in
the system and $n_{i\sigma}=c^{\dag}_{i\sigma}c_{i\sigma}$ is the
number operator. The parameters $t$ and $t'$ are nearest and
next-nearest neighbor hopping terms, respectively. The $L$-site ring system is
penetrated by magnetic flux $\Phi$, which is encapsulated in the hopping part of the Hamiltonian, such that,
\begin{equation}
\phi = \frac{2\pi}{L} \frac{\Phi}{\Phi_{0}}.
\end{equation}
The interaction term is the local Coulomb
interaction between opposite spins, the Hubbard $U$. All parameters are
specified in arbitrary units of energy and relative to the nearest-neighbor hopping parameter, $t$. Fig.~$1$ shows the geometry of the $t-t'-U$ ring with flux penetration. Note how the $t'/t$ interaction distorts the chain and introduces a quasi-$2$-dimensional aspect to the model. 

\begin{figure}[t]
\includegraphics[viewport= 15 6 572 434,height=5cm,width=5.7cm,clip]{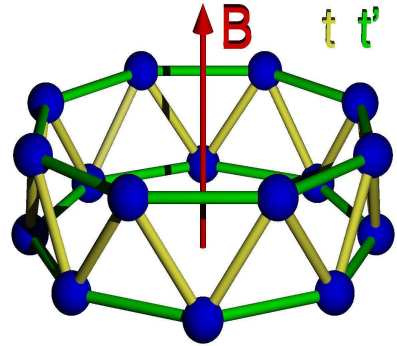}
\caption{Geometry of the $t-t'-U$ ring with penetrating magnetic
flux. In accordance with Ref.~\onlinecite{sd1}, the combination of
the nearest and next-nearest neighbor hopping terms is shown to
distort the $1D$ geometry, resulting in a system consisting two
coupled $1D$ chains. 
To illustrate this feature, the two rings are displaced along the direction of the magnetic
field. This is done for visualization purposes only, and
does not alter the physics of the model.}
\end{figure}

The basis functions for the lattice model are defined using an occupation number basis, 
\begin{equation}
\vert \psi_{\alpha}\rangle=\vert n_{\alpha 1\uparrow}...n_{\alpha N\uparrow};n_{\alpha 1\downarrow}...n_{\alpha N\downarrow}\rangle.
\end{equation}
The ground-state eigenstate $\vert \Psi\rangle_{lattice}$ is then a linear combination of these basis functions,  
\begin{equation}
\vert \Psi\rangle_{lattice}=\sum_{\alpha = 1}^{HS}a_{\alpha}\vert\psi_{\alpha}\rangle,
\end{equation}
where $a_{\alpha}$ are the coefficients of the basis states and $HS={L \choose N_{e\uparrow}}$ ${L \choose N_{e\downarrow}}$ is the size of the Hilbert space. Here, $N_{e\uparrow}$ and $N_{e\downarrow}$ refer to the number of spin-up and spin-down electrons, respectively. As only small system sizes are considered, the eigenproblem remains computationally tractable and can be solved by exact diagonalization using the ARPACK solver.\cite{ar1}
The ground-state energy $E_{0}$ and persistent current $I$ are investigated as a function of $\Phi$, where \begin{equation}
I(\Phi)=-\frac{\partial E_{0}(\Phi)}{\partial \Phi}. 
\end{equation}
To determine the magnetic state of the system, the total spin $S$ is calculated. The local moment, $\langle S_{i}^{2}\rangle=\frac{3}{4}\langle m^{2}_{i}\rangle$, is used to measure the degree of spin localization. For a quarter-filled chain the local moment has a maximum value of $\frac{3}{8}$.

\section{\label{sec:level1}RESULTS $\&$ DISCUSSION\protect\\ }
\subsection{Properties of the Flux-Penetrated $t-t'-U$ Chain as a Function of $U/t$ and $t'/t$}
Before discussing the physics of the flux-penetrated $t-t'-U$ ring, it will be necessary to review the flux penetrated Hubbard chain as this will provide an important base comparison for the $t-t'-U$ extended model. In Fig.~$2(a)$, the ground-state energy as a function of the flux is shown for an $8$-site Hubbard ring at quarter-filling. The results are for three different interaction strengths, $U/t=20$, $200$ and $1000$. The periodicity in the ground-state energy is found to change from $\Phi_{o}$ to $\Phi_{o}/N_{e}$ as the interaction strength is increased to large values and approaches perfect fractional periodicity as $U\rightarrow \infty$. At large interaction strengths, e.g., at $U/t=1000$, the angular momentum values at the minimum of the parabolas follow the expected $\Phi/\Phi_{o}=M/N_{e}$ relation. Thus the $M=0$, $1$, $2$, $3$ and $4$ angular momentum states, for example, correspond to $\Phi/\Phi_{o}= 0$, $0.25$, $0.5$, $0.75$ and $1.0$ respectively.  

In Fig.~$2(b)$, the persistent current is shown for the $U/t=1000$ Hubbard ring.  The change in positive to negative current corresponds to the minima of the parabolas in Fig.~$2(a)$. The jump from minimum to maximum current is indicative of the change in angular momentum state of this system. The amplitude of the persistent current is seen to be almost constant as a function of the penetrating flux. This near-perfect fractional periodicity corresponds to the system being an ``effective rigid-rotator".  At $U=\infty$, perfect fractional periodicity and, hence, full rigid rotation would occur. In this case, the amplitude of the persistent current would be constant. 

\begin{figure}[t]
\resizebox {8cm}{6.5cm}{
\includegraphics{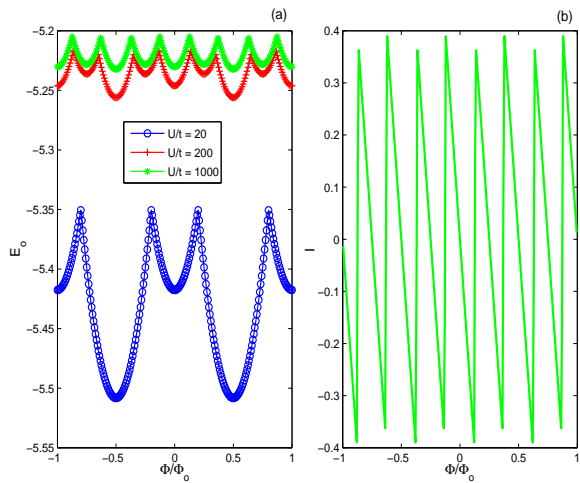}}
\caption{(a) Ground-state energy as a function of flux for an $8$-site Hubbard system at quarter-filling $(2\uparrow,2\downarrow)$, $U/t=20$, $U/t=200$ and $1000$. (b) Persistent current as a function of flux for the $U/t=1000$ result.} 
\end{figure}

\begin{figure}[t]
\resizebox {8cm}{6.5cm}{
\includegraphics{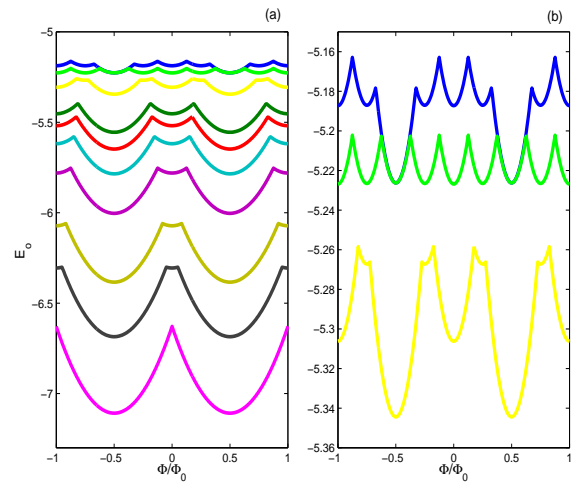}}
\caption{(a) Ground-state energy as a function of flux for an $8$-site $t-t'-U$ system at quarter-filling $(2\uparrow,2\downarrow)$, with $t'/t=-0.05$. The values of $U/t$ starting from the bottom curve are $0$, $1$, $2$, $4$, $6$, $8$, $10$, $20$, $40.2$ and $1000$. $U/t=40.2$ is specifically shown as this indicates the onset of effective rigid rotation, and hence near perfect fractional periodicity. (b) Ground-state energy as a function of flux, with focus on the $U/t=20$, $40.2$ and $1000$ results.} 
\end{figure}

Fig.~$3$ shows the ground-state energy for the quarter-filled $t-t'-U$ system as a function of flux, with $U/t=8$ and $t'/t=-0.05$.  The application of the small $t'/t$ term is found to induce marked differences in the ground-state energy as a function of the flux strength, thus, demonstrating the sensitivity of the solutions to small $t'/t$ perturbation. An interesting deviation from the Hubbard results occurs---namely, an early onset of the fractional periodicity with respect to $U/t$. This effect is clearly evidenced in Fig.~$3(b)$ by the appearance of the $\Phi/\Phi_{o}=\pm0.25$ parabolas in the $U/t=20$ result. At $U/t\sim 40.2$, the $t-t'-U$ system becomes an effective rigid-rotator, which can be seen in the near-perfect fractional periodicity of the $U/t = 40.2$ solution. In comparison, for the Hubbard ring, larger values of the Hubbard $U$ (i.e., $U/t\geq1000$) were required in order to achieve this effect. The Hubbard results show that the fractional periodicity improves as a function of increasing Hubbard $U$ strength. In the $t-t'-U$ system, however, the fractional periodicity acquires its optimal state at moderate values of the Hubbard $U$. Increasing $U/t$ for the $t-t'-U$ system destroys the near-perfect fractional periodicity due to the `freezing' of the energetics in the $\Phi/\Phi_{o}=\pm0.5$ result as a function of increasing $U/t$. In this case, the energy corresponding to the ground-state solution at $\Phi/\Phi_{o}=\pm0.5$ remains the same for $U/t\ge 40.2$. 

Inspection of the total spin as a function of $U/t$ at
$\Phi/\Phi_{o}=\pm0.5$ shows that the system has undergone a magnetic
transition at $U/t\sim 40.2$, from a $S=0$ to a $S=2$ fully-polarized
ferromagnetic spin state. The ferromagnetic transition can be
confirmed by comparing the energies corresponding to the lowest energy
$S=0$ and $S=2$ spin states at $\Phi/\Phi_{o}=\pm0.5$, as a function
of increasing $U/t$.  At $U/t=20$, the ground-state has $S=0$ with an
energy gap of 0.12 to the next state, which is $S=2$. At $U/t\sim
40.2$, the lowest energy $S=0$ and $S=2$ spin states are near
degenerate---the $S=2$ ferromagnetic spin state being the ground-state
solution. Away from this point at $U/t=1000$, the ferromagnetic $S=2$
ground-state persists, with the magnitude of the energy difference
between the lowest energy $S=0$ and $S=2$ spin states, again, being
$0.12$. The freezing of the energetics at $\Phi/\Phi_{o}=\pm0.5$ is
therefore related to the removal of Hubbard $U$ effects in the
fully-polarized (ferromagnetic) state---in this case, no double
occupancy and hence Hubbard $U$ contribution can occur. The onset of
effective rigid rotation and near-perfect fractional periodicity
corresponds to a system where the spins are well localized. Thus, the
local moments for the $S=2$ ferromagnetic spin-state at
$\Phi/\Phi_{o}=\pm0.5$ reach the maximum value of $\langle
S^{2}_{i}\rangle=0.375$. At $U/t=40.2$, the local moment results are
$0.3748$ for $\Phi/\Phi_{o}=0$ and $\Phi/\Phi_{o}=\pm0.25$. These
results are close to the maximum value of $\langle
S^{2}_{i}\rangle=0.375$. At $U/t\ge 40.2$ the system continues to
evolve as a function of $U/t$ for all states other than
$\Phi/\Phi_{o}=\pm0.5$. Fig.~$4$ summarizes the local moment results
for the $t-t'-U$ system at $\Phi/\Phi_{o}=0.5$, for the $t'/t=0$
(Hubbard) and $t'/t=-0.05$ cases. The ferromagnetic transition for the
$t'/t=-0.05$ system is indicated by a change in spin-state to the
fully-polarized $S=2$ state, together with a corresponding jump in the
local moment to the maximum value of $0.375$. Compared to this result,
the $t'/t=0$ (Hubbard) solutions show a smooth evolution of the
$\Phi/\Phi_{o}=0.5$ state with respect to the local moment solutions
and no change in the $S=0$ spin-state of the system.

\begin{figure}[t]
\resizebox {8cm}{6.5cm}{
\includegraphics{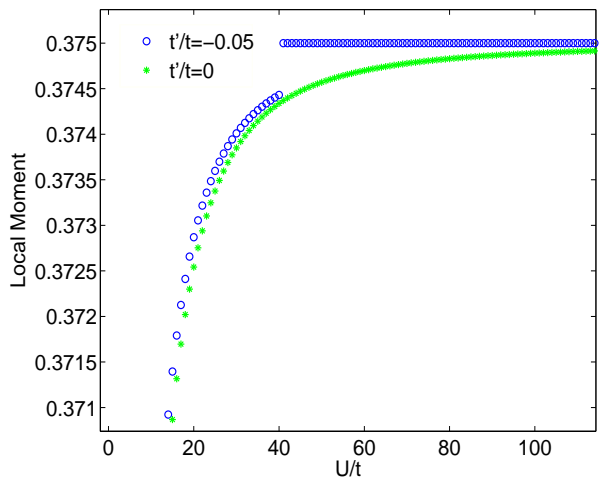}}
\caption{(a) The local moment $\langle S^{2}_{i}\rangle$ as a function of $U/t$ for an $8$-site $t-t'-U$ system at quarter-filling, $(2\uparrow,2\downarrow)$. The two solutions are the  $\Phi/\Phi_{o}=0.5$, $t'/t=0$ and $t'/t=-0.05$ results, as indicated in the legend. Note the discontinuity in the $t'/t=-0.05$ solution, which occurs at the onset of effective rigid rotation ($U/t\simeq 40.2$) and corresponds to the $S=0\rightarrow 2$ ferromagnetic transition.} 
\end{figure}

Fig.~$5$ shows how the effective rigid-rotator feature of the $t-t'-U$ system evolves as a function of $t'/t$. Increasing the magnitude of $t'/t$ to $0.1$ decreases the effective rigid rotation value of $U/t$ to $U/t\sim 16.9$. The onset of effective rigid rotation again corresponds to a ferromagnetic transition to the fully-polarized $(S=2)$ state and freezing of the energetics in the $\Phi/\Phi_{o}=\pm0.5$ results. Consequently, the local moment solutions at $\Phi/\Phi_{o}=\pm0.5$ with $U/t= 16.9$ have maximum values of $0.375$, with local moment results for the $\Phi/\Phi_{o}=0$ and $\pm0.25$ solutions being $0.3741$ in these cases. 

In general, calculations for the $t-t'-U$ system as a function of $t'/t$ reveal an inverse relation between $t'/t$ and $U/t$ with respect to the onset of near-perfect fractional periodicity and effective rigid rotation. Table I shows the point of onset of effective rigid rotation as a function of $t'/t$ and $U/t$. Note, this onset also corresponds to the ferromagnetic $S=0\rightarrow 2$ transition in the $\Phi/\Phi_{o}=0.5$ result. The inverse relation between the $t'/t$ and $U/t$ parameters seen in Table I and corresponding ferromagnetic transition are in accordance with the known properties of the quarter-filled $t-t'-U$ chain---namely that the critcal parameter setting for ferromagnetism occurs when $U \sim \vert t'\vert^{-1}$---see Ref.~\onlinecite{sd1}. 

\begin{figure}[t]
\resizebox {8cm}{6.5cm}{
\includegraphics{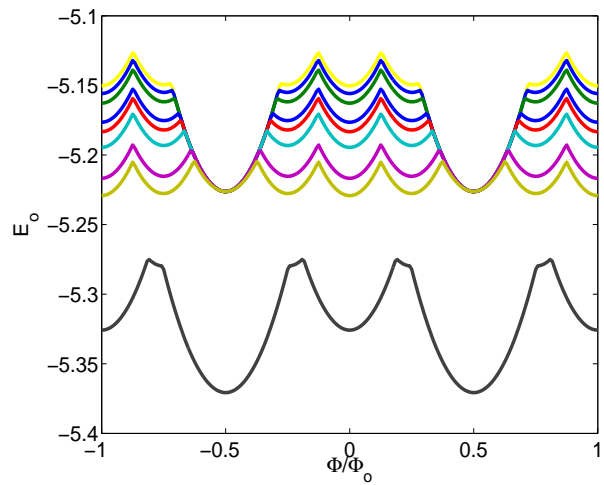}}
\caption{(a) Ground-state energy as a function of flux for an $8$-site $t-t'-U$ system at quarter-filling, $(2\uparrow,2\downarrow)$, with $t'/t=-0.1$. The values of $U/t$ starting from the bottom curve are $10$, $16.9$, $20$, $30$, $40$, $50$, $100$, $200$, and $1000$.}
\end{figure}

\begin{table}
\caption {Parameter values corresponding to the onset of effective rigid rotation, $t'/t$ and $U/t$, together with $\Delta E_{max}$ and the local moment results, $\langle S^{2}_{i}\rangle$, for the $8$-site quarter-filled $t-t'-U$ system. The first result is for the Hubbard model ($t'=0$). For $t'\ne 0$, the $U/t$ values also indicate the point of transition to the ferromagnetic $S=2$ state, which occurs at $\Phi/\Phi_{o}=0.5$.}
\begin{ruledtabular}
\begin{tabular}{cccccc}
$t'/t$&$\sim U/t$&$\Delta E_{max}$&$\langle S^{2}_{i}\rangle_{\Phi/\Phi_{o}=0}$&$\langle S^{2}_{i}\rangle_{\Phi/\Phi_{o}=0.25}$&$\langle S^{2}_{i}\rangle_{\Phi/\Phi_{o}=0.5}$\\
\hline
\hline
\\
$0$ &$1000$&$0.004$&$0.3750$&$0.3750$&$0.3750$\\
$-0.05$ &$40.2$&$0.001$&$0.3748$&$0.3748$&$0.3750$\\
$-0.1$ & $16.9$&$0.003$&$0.3741$&$0.3741$&$0.3750$\\
$-0.2$ & $5.3$&$0.020$&$0.3701$&$0.3701$&$0.3750$\\
$-0.5$ & $3.0$&$ 0.076$&$0.3671$&$0.3681$&$0.3750$\\
\end{tabular}
\end{ruledtabular}
\end{table}

The $\Delta E_{max}$ values in Table I are defined as the maximum energy difference between the minima of the parabolas at $\Phi/\Phi_{o}=0$, $0.25$ and $0.5$. At low values of $t'/t$, the corresponding $\Delta E_{max}$ values are small, meaning that the system has near-perfect fractional periodicity. At $t'/t=-0.05$ and $-0.1$, for example, the values for $\Delta E_{max}$ are $0.001$ and $0.003$ respectively. The relative closeness in energy in these solutions can be explained by a minimum Hubbard $U$ contribution to the ground-state results occurring at $\Phi/\Phi_{o}=0$, $0.25$ and $0.5$. Inspection of the highest weighted basis states in the ground-state wavefunctions show that at the point of effective rigid rotation there are no double occupancies, and, hence, no significant Hubbard $U$ contribution to the ground-state results. Increasing $\vert t'/t \vert$ adds addtional $t'/t$ kinetics thereby causing an increase in $\Delta E_{max}$. This is also reflected in Table I in the corresponding reduction in the local moment results. 

In summary, effective rigid rotation in the $t-t'-U$ ring can be defined by the following three criteria, 

$(1)$ near-perfect $\Phi_{o}/N_{e}$ fractional periodicity, with energy minima occurring at values of $\Phi/\Phi_{o}=M/N_{e}$. These features are characterized by the relative closeness in the energy minima of the parabolas as defined by $\Delta E_{max}$ and can be accounted for by the freezing out of the $U/t$ energetics at $\Phi/\Phi_{o}=\pm0.5$, together with a minimum Hubbard $U$ contribution at other  $\Phi/\Phi_{o}=M/N_{e}$ values. $(2)$ An $S=0\rightarrow 2$ ferromagnetic transition, which occurs at $\Phi/\Phi_{o}=\pm0.5$, and $(3)$ maximum or near-maximum values in the local moments determined at these points. 

The following trends can also be seen in the persistent current results.  Comparison of the persistent current for the Hubbard (Fig.~$6(a)$) and $t-t'-U$ (Fig.~$6(b)$) systems show an increase in the number of peaks found in the $I(\Phi)$ curve. In particular, the $U/t=20$, $t-t'-U$ solution shows double the number of peaks compared to the Hubbard solution. Early onset of effective rigid rotation is also evident, and this is characterized by equal amplitude peaks in the $I(\Phi)$ function. Of significance are the similarities in the results for the Hubbard system at $U/t=1000$ (Fig.~$6(a)$) and the $t-t'-U$ system at $U/t=40.2$ (Fig.~$6(b)$)---both showing effective rigid rotation. Characteristic of the $t-t'-U$ system is that it undergoes a ferromagnetic transition leading to the freezing of the energetics at $\Phi/\Phi_{o}=\pm 0.5$. In addition, at $\Phi/\Phi_{o}\ne\pm 0.5$ the energetics continue to evolve beyond the point of effective rigid rotation. This leads to a distinctive feature in the persistent current curve, namely in the formation of signal `clusters' consisting of the same number of peaks as there are particles in the system. An example of such a cluster is shown in Fig.~$6$(b) for $U/t=1000$ at $-0.5\le \Phi/\Phi_{o}\le 0.5$. 

\begin{figure}[t]
\resizebox {8cm}{6.5cm}{
\includegraphics{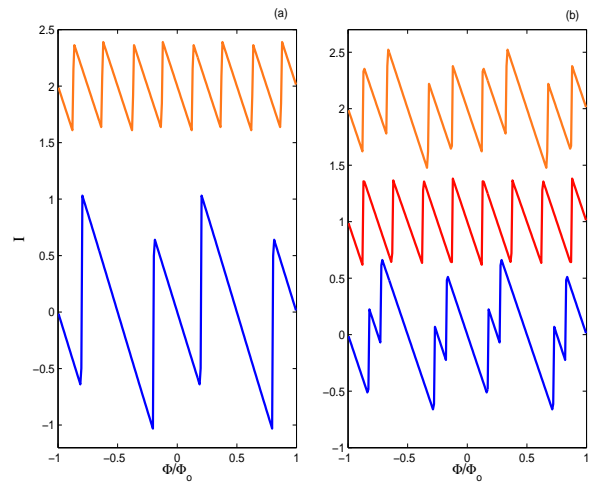}}
\caption{Persistent current as a function of flux, for the $8$-site, quarter-filled (a) Hubbard chain with $U/t=20$ and $1000$ and (b) $t-t'-U$ system at $t'/t=-0.05$ and $U/t=20$, $40.2$ and $1000$ respectively. Note that the $I=0$ axis has been shifted in order to separate the results for comparison purposes.} 
\end{figure}

\subsection{Correspondence Between the Flux Penetrated $t-t'-U$ Ring and Few-Particle 2D Hard-Wall Quantum Dot}
A circularly symmetric few-particle hard-wall quantum dot (QD) is an example of an extended quasi-ring-like system. 
Such structures can be formed in semiconductor heterostructures by etching techniques or
electrical gating of the two-dimensional electron gas (2DEG).
In a typical example of the ${\rm GaAs}/{\rm Al}_{x}{\rm
Ga}_{1-x}{\rm As}$ heterostructure, the trapped electrons
can be modelled as two-dimensional droplets of
charges with renormalized mass, $m^{*}= 0.067 m_e$, and dielectric constant,
$\epsilon = 12.7\epsilon_0$. 

A conventional choice for a model Hamiltonian for this system is
\begin{equation}
H=\sum^N_{i=1}\left[
 \frac{({\bf p}_i+e {\bf A} )^2}{2 m^*}
+V(r_i)\right] + \frac{e^2}{4\pi \epsilon} \sum_{i<j}
\frac{C}{r_{ij}},
\label{eq:hamiltonian}
\end{equation}
where $V(r)$ is the external confinement potential.
In the low-density regime and at high magnetic field, the interactions
between the charges are significant. While a realistic interaction 
potential is much softer due to screening effects, a long-range
Coulomb interaction whose strength is parameterized by the dimensionless
scaling factor $C$ will be assumed. Additionally, an homogeneous external magnetic
field ${\mbf B} = \nabla\times {\mbf A}$, perpendicular to the 2DEG plane will be included.
The confinement potential is taken to be of the form
\begin{eqnarray}
V(r)&=& 0, \ r\le R \nonumber\\
&=& \infty, \ r > R,
\end{eqnarray}
where $R=5a_0^*$ is the radius of the dot and $a_0^*$ is the effective
Bohr radius.  The choice of the hard-wall potential is motivated by
the aim of having extended ring-shaped electronic structures with
ferromagnetic ground-state solutions. This potential has been used in
Ref.~\onlinecite{fg1} to study the effect of confinement on the
magneto-optical spectrum of a QD, and in Refs. \onlinecite{esa1} and
\onlinecite{esa2} to study the electronic structure of a QD in a
strong magnetic field, using a mean-field density-functional
method. The most common choice of potential for quantum dots, the
parabolic potential, is not a feasible choice as the ground-states are
not ring-shaped in the regime of interest.
In addition, the ground-states are not ferromagnetic without a strong
Zeeman term, despite the effects of Landau level mixing \cite{ss1}. This can be seen, e.g., in Ref.~\onlinecite{kos1}, which
shows exact diagonalization results for electrons in a parabolic
confinement, treated in the lowest-Landau-level approximation.
The Zeeman coupling is therefore neglected in this work to better elucidate the
interesting fact that the spin-polarized solutions are caused by the
electron-electron interaction and not by the Zeeman term.

The ground-state properties of the QD system are determined numerically by diagonalizing
the Hamiltonian in Eq. (\ref{eq:hamiltonian}) in a basis of spin-dependent
Fock states \cite{sigga} built from a truncated set of
lowest-lying single-particle eigenstates of the system.\cite{fg1} The z-component of the spin, $S_{z}$, and the angular momentum, $M$, are found to be good quantum numbers for this basis. The 
functional form of the energy eigenstates can be obtained analytically,\cite{fg1} however, this has explicit dependence on the respective energy eigenvalue, 
which has to be solved numerically. The matrix elements of the
Coulomb interactions for pairs of single-particle states are also 
evaluated numerically.

\begin{figure}[t]
\begin{center}
  \includegraphics[width=.9\columnwidth, height=7cm]{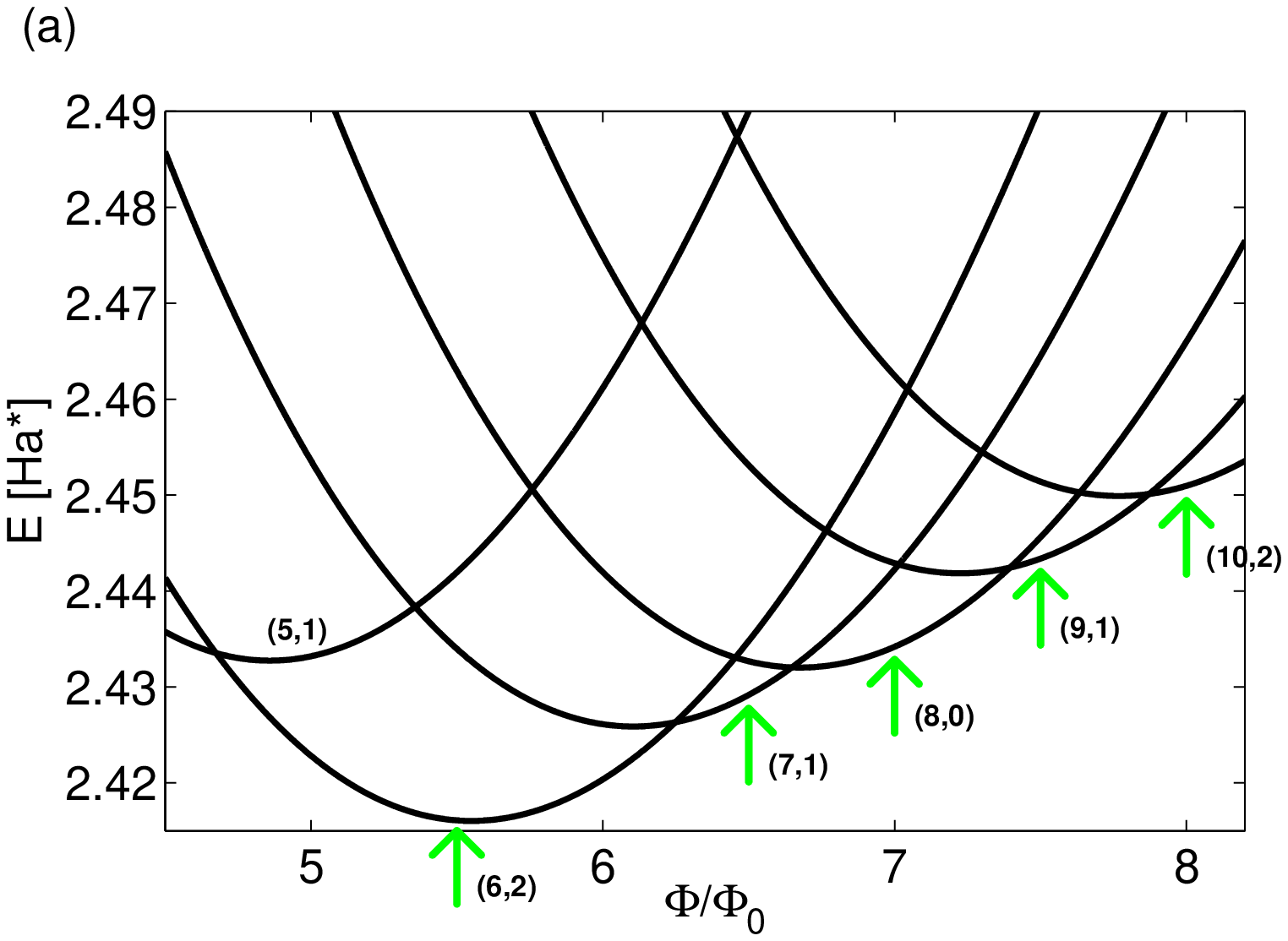}
  \includegraphics[width=.9\columnwidth, height=7cm]{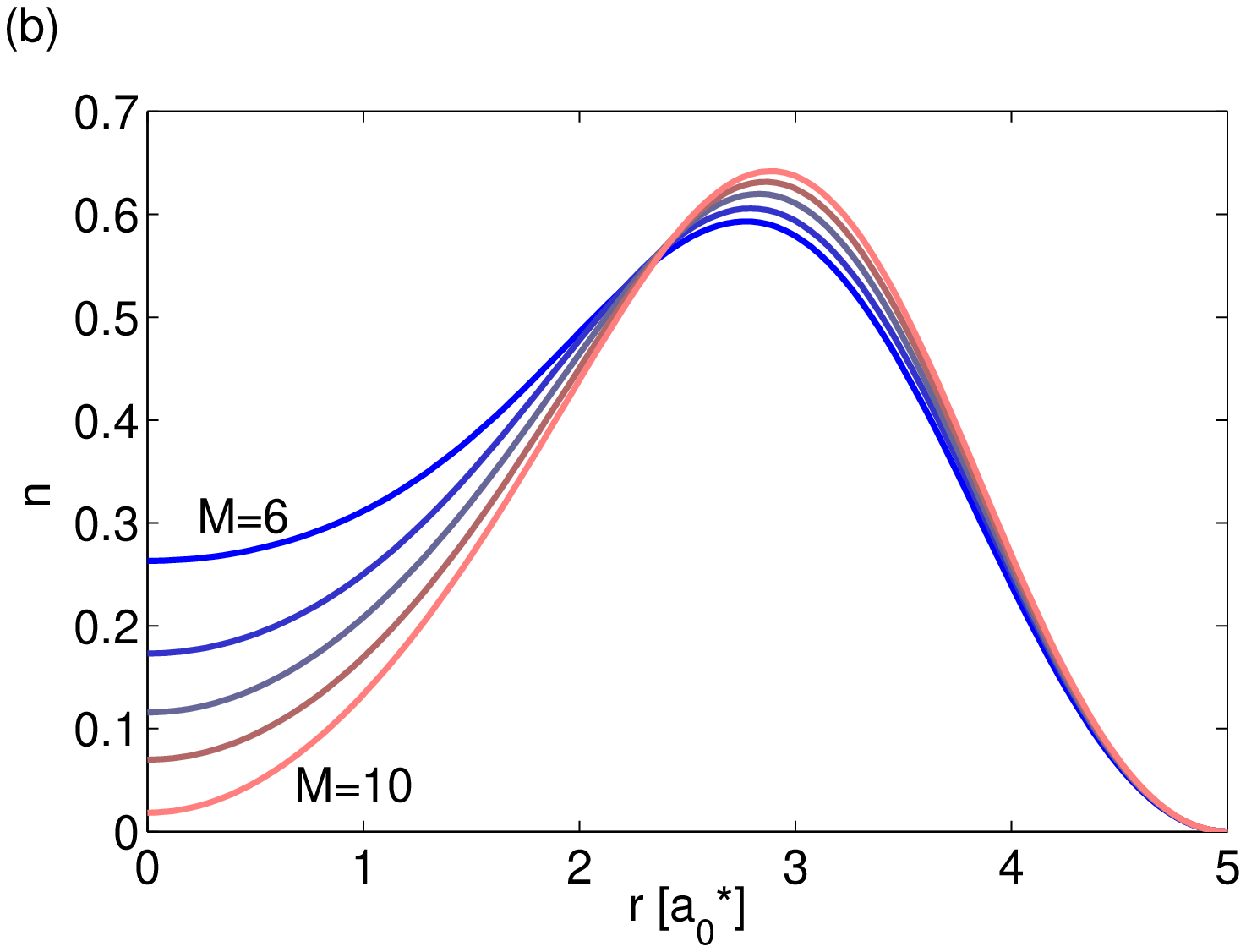}
  \caption{(a) Ground-state energies and (b) radial electron densities
    for the $4$-particle hard-wall quantum dot, with $M=6,\ldots,10$. The quantum numbers of the levels are shown in (a) as the pair $(M,S)$. The cyclotron energy, $\hbar\omega_c/2$, of the zero-point motion has been substracted from these graphs. }
  \label{fig:Fig7}
\end{center}
\end{figure}

Fig.~\ref{fig:Fig7} shows the ground-state energy results and the
radial electron densities obtained from the eigensolutions of
Eq. (\ref{eq:hamiltonian}), with corresponding $(M,S)$ values. In
Fig.~\ref{fig:Fig7}(a) the angular momentum of the ground-state energy
solutions can be seen to increase as a function of the increasing
total flux, $\Phi = \pi |{\mbf B}|R^2$, in the dot. The electron
density distribution (Fig.~\ref{fig:Fig7}(b)) assumes a
quasi-ring-like behavior, with the maximum density occurring near the
mid-radius. As the flux through the dot and hence the angular momentum
is increased, the electron distribution forms a tighter ring-like
structure. For example, at $M=10$, the radial electron density tends
to zero at $r=0$ and at the edge of the dot at $r=5a_{0}$. 

\begin{figure}[t]
\begin{center}
  \includegraphics[width=8.5cm, height=8cm]{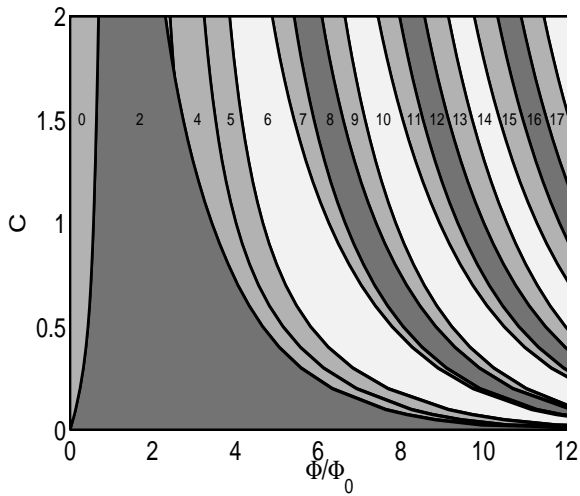}
  \caption{Ground-state phase diagram of a $4$-particle hard-wall quantum dot with $R = 5a_0^*$
    as a function of the external magnetic flux, $\Phi$, and effective interaction strength, $C$.
    The shade of the filling
    represents the spin quantum number, from dark to light $S= 0,1,2$, and
    the integers represent the angular momentum $M$ of the state.
    Fractional quasi-periodicity appears at strong $\Phi$ and $C$.}
  \label{fig:Fig8} 
\end{center}
\end{figure}

Further evidence of the quasi-ring-like behavior can be seen in
Fig.~\ref{fig:Fig8}, which shows the ground-state phase diagram for
the dot as a function of $\Phi$ and $C$.  Fractional quasi-periodicity
in the ground-state parameters occurs as a function of the flux for
strong values of $\Phi$ and $C$. Consistent with the flux-penetrated
1D Hubbard ring, the ground-state total spin varies periodically, with
the angular momentum increasing in minimal steps of one. The spin-polarized 
ground-states appear in angular momentum steps of $N_{e}$, which is similar to parabolic QD's in the regime of small electron numbers. \cite{ss2}

Interestingly, the total spin oscillation as a function of $M$ for the hard-wall QD (Fig.~$8$) is not the same as that reported by Koskinen $\it{et}$ ${\it al.}$ for the quasi-one-dimensional rings \cite{mk1}---the latter system corresponds to the flux-penetrated anti-ferromagnetic Heisenberg model, whereas the hard-wall QD corresponds to the flux-penetrated ferromagnetic Heisenberg model. An important question then arises as to what are the
microscopic mechanisms for these differences between the $1D$ ring and
the hard-wall $2$D systems?

In order to address this question, the continuum model solutions as a function of $M$ and $S$ are compared with those from the $t-t'-U$ lattice model, considered to be the minimal itinerant model for ferromagnetism in $1D$ systems. For convenience, the continuum model results are chosen at the flux values shown by the arrows in Fig.~\ref{fig:Fig7}(a).
Since the lattice basis is not complete in the continuum 
description, a direct comparison is not possible. Instead, the many-body wavefunction 
$\Psi({\mbf x}_1,\ldots,{\mbf x}_N)$ is projected onto a tensor-product basis of delta functions
\begin{equation}
  |j,\sigma_z\rangle = \delta({\mbf x}-{\mbf x}_j^*)\delta_{\sigma,\sigma_z},
\end{equation}
representative of the lattice basis, with the
resulting Fock expansion normalised to unity. Here, the lattice index $j$ runs over the lattice sites,
and ${\mbf x}^*_j$ is the real-space coordinate corresponding to the $j$th site.
The choice of ${\mbf x}^*_j$ is not unique, as the only constraint is that
the sites should form a ring-like lattice. To obtain a unique lattice, 
the radius is chosen to maximize the amplitude of the many-body wavefunction
with its arguments assuming a regular polygonal configuration.
Once the wavefunction is expressed in the lattice basis, 
the overlap between the continuum and 
corresponding lattice eigenstate solutions, $\langle \Psi_{cont}| \Psi_{lattice}\rangle$, can then be calculated.

One of the important aspects of this work is to understand the role of
the electronic itineracy with respect to the magnetism seen in the
continuum model results. To determine the relationship between these
properties, the $N_{e}$ particle continuum model solutions have been
projected onto a discrete quarter-filled lattice model, having $(2$ x
$N_{e})$ lattice sites. To justify this choice of lattice, a specific
example is given of the $4$-particle continuum model results projected
onto an $8$-site, quarter-filled lattice. Here the maximum density
droplet (MDD) state is considered, which refers to the lowest angular
momentum state of the QD when the electrons are in the lowest Landau level.\cite{ahm1} The angular momentum of the MDD state is defined as
$M_{\mathrm{MDD}}=N_{e}(N_{e}-1)/2$. For the $N_{e}=4$ particle QD
system, $M_{\mathrm{MDD}}=6$. Inspection of the normalized
ground-state set of eigenvector coefficients determined from the
continuum model projection onto the discrete lattice demonstrates that
the most highly weighted basis states for the $M_{\mathrm{MDD}}=6$
ground-state are those which are of Heisenberg-type. For the
quarter-filled, 8-site chain with $S_{z}=0$, these highly weighted
states correspond to those of type $(\uparrow 0 \downarrow 0 \uparrow
0 \downarrow 0)$ and $(\uparrow 0 \uparrow 0 \downarrow 0 \downarrow
0)$, with perturbations thereof. There are $12$ states of this type
within the Hilbert space, which consists of a total of $784$ basis
states. For the normalized wavefunction, the corresponding weightings
for each of these states is equal to $0.018$. The second highest
weightings are associated with basis states of type $(\uparrow
\uparrow 0 \downarrow 0 \downarrow 0 0)$ and $(\uparrow \downarrow 0
\uparrow 0 \downarrow 0 0)$, together with perturbations of these,
adding up to $96$ basis states in total. These states have associated
weightings of $0.005$ each. Note specifically that the second highest
weighted states have nearest-neighbor spins. This justifies the use of
the extended, quarter-filled lattice and brings to question the role
of the kinetics with respect to the ferromagnetism seen in the
continuum model result.

\begin{table}
\caption {Overlap results between the $8$-site, quarter-filled $t-t'-U$ ring and $4$-particle continuum model system. The lattice model parameters are $U/t=23$ and $t'/t=-0.17$.}
\begin{ruledtabular}
\begin{tabular}{cccc}
$L_{cont}$&($\Phi/\Phi_{0})_{cont}$&($\Phi/\Phi_{0})_{lattice}$&$\%$ $Overlap$\\
\hline
\hline
\\
$6$ &$5.5$&$1.5$&$98.4$ \\
$7$ &$6.5$&$1.75$&$98.2$  \\
$8$ &$7.0$&$2.0$&$97.8$  \\
$9$ &$7.5$&$2.25$&$97.5$  \\
$10$ &$8.0$&$2.5$&$97.1$ \\
\end{tabular}
\end{ruledtabular}
\end{table}

\begin{table}
\caption {Overlap results between the $10$-site, quarter-filled $t-t'-U$ ring and $5$-particle continuum model system. The lattice model parameters are $U/t=13$ and $t'/t=-0.19$.}
\begin{ruledtabular}
\begin{tabular}{cccc}
$L_{cont}$&($\Phi/\Phi_{0})_{cont}$&($\Phi/\Phi_{0})_{lattice}$&$\%$ $Overlap$\\
\hline
\hline
\\
$10$ &$6.9$&$2.0$&$98.5$ \\
$11$ &$7.3$&$2.2$&$98.1$  \\
$12$ &$7.7$&$2.4$&$97.8$  \\
$13$ &$8.1$&$2.6$&$97.7$  \\
$14$ &$8.5$&$2.8$&$97.6$ \\
$15$ &$8.8$&$3.0$&$97.4$ \\
\end{tabular}
\end{ruledtabular}
\end{table}

To determine the correspondence between the $t-t'-U$ model and continuum model ground-state wavefunctions, the overlaps, $\langle \Psi_{cont}| \Psi_{lattice}\rangle$, have been calculated. Both the $4$-particle, $S_{z}=0$, and $5$-particle, $S_{z}=1/2$, continuum model cases have been considered. These cases were compared to quarter-filled $8$- and $10$-site $t-t'-U$ systems, respectively. The results for the maximum overlaps between the continuum and lattice model results are shown in Tables II and III. Here, the $(\Phi/\Phi_{0})_{cont}$ values denote the total amount of magnetic flux that is penetrating the dot. For the 4-particle case, these values correspond to the flux values indicated in Fig.~\ref{fig:Fig7}(a). A similar choice of flux values was made for the $5$-particle system. The maximum overlaps were determined for the first period in $M$, commencing at the MDD states, $M_{MDD}=6$ and $M_{MDD}=10$, for the $4$- and $5$-particle systems, respectively. The results were obtained by careful scanning through the $U/t$ and $t'/t$ parameter ranges. 

Maximum overlaps as a function of $M$ for the $4$-particle system were found to occur at $U/t=23$ and $t'/t=-0.17$ and for the $5$-particle system at $U/t=13$ and $t'/t=-0.19$. The dominant difference in these results lies in the strength of the Hubbard $U$. A larger $U/t$ value for the 4-particle system is consistent with what is known in the continuum model description of quantum rings, namely that these systems become more interaction dominated when they contain fewer particles. Given the similar magnitude of $t'/t$ in both the $4$-particle and $5$-particle cases, it seems apparent that the role of this parameter is therefore to simulate the correct extended geometry of the ring. As $M$ is increased, the continuum model systems become more ring-like, losing some of these extended system qualities. Thus, for larger $M$, the maximum overlaps for both the $4$- and $5$-particles cases are found to slightly decrease. 

\begin{figure}[t]
\resizebox {8cm}{6.5cm}{
\includegraphics{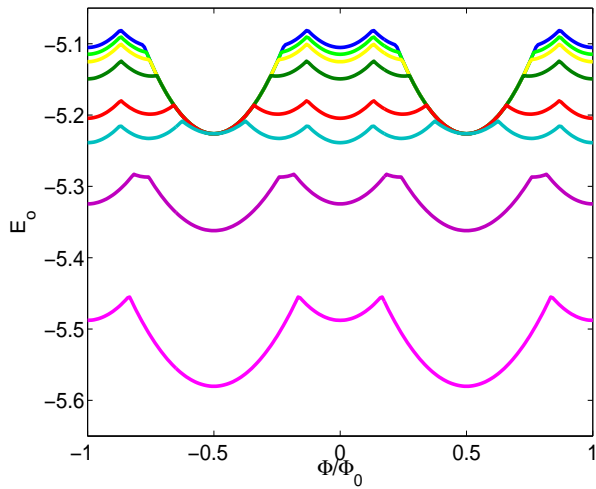}}
\caption{Ground-state energy as a function of flux for an 8-site, quarter-filled $t-t'-U$ chain with $t'/t=-0.17$. The values of $U/t$ starting from the bottom curve are $3$, $5$, $7.28$, $10$, $23$, $50$, $100$ and $1000$ respectively. Note the point of effective rigid rotation occurs at $\simeq U/t=7.28$.} 
\end{figure}

In Figs.~$9$ and $10$ the ground-state energy results as a function of flux for the 8-site and 10-site $t-t'-U$ ring systems are shown with the choice of $t'/t$ values being the same as those which were required to achieve maximum overlap. The values of $U/t$ required for maximum overlap occur after the point of effective rigid rotation in both of these cases. From the lattice model perspective, this result shows that the system has undergone the ferromagnetic transition required to initiate effective rigid rotation. The total spin progression for both the continuum and lattice model systems as a function of $M$ therefore follows that of the flux-penetrated ferromagnetic Heisenberg model, namely, $(M,S)_{4}=(6,2)\rightarrow (7,1)\rightarrow (8,0)\rightarrow (9,1)\rightarrow (10,2)$ and $(M,S)_{5}=(10,\frac{5}{2})\rightarrow (11,\frac{3}{2})\rightarrow (12,\frac{1}{2})\rightarrow (13,\frac{1}{2})\rightarrow (14,\frac{3}{2}) \rightarrow (15,\frac{5}{2}) $, for the 4-particle and 5-particle cases. The high correspondence between the continuum and lattice model systems, being between $97-98\%$, demonstrates clearly that, similar to the $t-t'-U$ system, the additional degree of kinetic freedom and extended ring-like geometry are the important factors which are responsible for the ferromagnetism seen in the continuum model results. 

\begin{figure}[t]
\resizebox {8cm}{6.5cm}{
\includegraphics{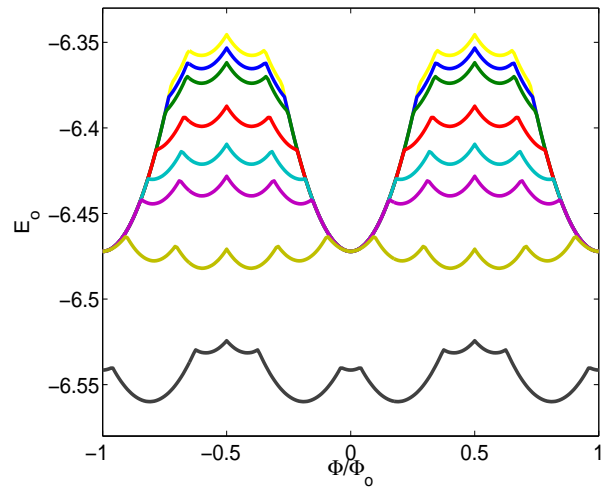}}
\caption{Ground-state energy as a function of flux for the 10-site, quarter-filled $t-t'-U$ chain with $t'/t=-0.19$. The values of $U/t$ starting from the bottom curve are $5$, $6.47$, $10$, $13$, $20$, $50$, $100$ and $1000$ respectively. Note the point of effective rigid rotation occurs at $\simeq U/t=6.47$.} 
\end{figure}

In summary, to encapsulate the correct physics and the kinetics in the continuum QD system, the $t-t'-U$ lattice needs to be double in size and hence quarter-filled with respect to the total particle number. To obtain maximum overlap, the lattice model needs to have evolved beyond the point of rigid rotation, which is denoted by the onset of a ferromagnetic phase, thus corresponding to the ferromagnetism seen in the continuum model results. 

\section{\label{sec:level1}CONCLUSION\protect\\}
The flux-penetrated $t-t'-U$, $1D$ system at quarter-filling has been investigated as a function of $t'/t$ and $U/t$ parameterization. The results indicate an onset of effective rigid rotation in this system, which occurs for moderate $U/t$ values, and coincides with a ferromagnetic transition. This model has been used to explain the essential physics of the continuum model results for a few particle, $2D$ quantum dot. For $4$- and $5$-particle cases, the maximum overlaps between the continuum and lattice model wavefunctions give between $97-98\%$ correspondence. The ferromagnetism, which is seen in the continuum model solutions, can be explained by the additional kinetics and extended ring-like geometry. The results suggest the possibility of anti-ferromagnetic to ferromagnetic switching from strictly $1D$ to extended quantum ring systems. In addition, a decrease in the order of magnitude of the Hilbert space size in the lattice model compared to the continuum model indicates that the lattice model may be a more efficient method of determining the ground-state properties and essential physics of extended ring QD systems.

\section{ACKNOWLEDGEMENTS}
This work was supported by the Academy of Finland through its Centre of Excellence Program (2000--2011).

\end{document}